\documentclass[10pt,twocolumn]{article}

\usepackage[a4paper,margin=2cm,columnsep=0.7cm]{geometry}
\usepackage[T1]{fontenc}
\usepackage{graphicx}
\usepackage{amsmath}
\usepackage{siunitx}
\usepackage[version=4]{mhchem}
\usepackage{booktabs}
\usepackage{authblk}
\usepackage{graphicx}
\usepackage{float}
\usepackage[colorlinks=true,citecolor=blue,linkcolor=blue,urlcolor=blue]{hyperref}

\usepackage[super,sort&compress]{natbib}

\newcommand{\affiliation}[2][]{\affil[#1]{#2}}

\title{Ultrafast Dissipative Localization of Electronic Energy in Au--Pt Superlattices}

\author[1]{C.~Walz}
\author[2,*]{M.~Mattern}
\author[3]{J.-E.~Pudell}
\author[2]{J.~Jarecki}
\author[1]{L.~Mehner}
\author[1,4]{F.-C.~Weber}
\author[1]{F.~Baltrusch}
\author[1]{S.~P.~Zeuschner}
\author[1]{M.~Herzog}
\author[5]{B.~Ahn}
\author[5]{J.~C.~Ekstr\"om}
\author[5]{D.~Kroon}
\author[5]{A.~Jurgilaitis}
\author[5]{J.~Larsson}
\author[6]{M.~Kronseder}
\author[2]{D.~Schick}
\author[1]{A.~von Reppert}
\author[1,4,*]{M.~Bargheer}

\affiliation[1]{Institut f\"ur Physik \& Astronomie, Universit\"at Potsdam, 14476 Potsdam, Germany}
\affiliation[2]{Max Born Institute for Nonlinear Optics and Short Pulse Spectroscopy, 12489 Berlin, Germany}
\affiliation[3]{European X-ray Free-Electron Laser Facility, 22869 Schenefeld, Germany}
\affiliation[4]{Helmholtz-Zentrum Berlin, 12489 Berlin, Germany}
\affiliation[5]{MAX IV Laboratory, Lund University, PO Box 118, SE-221 00 Lund, Sweden}
\affiliation[6]{Institute of Experimental and Applied Physics, University of Regensburg, 93053 Regensburg, Germany}

\date{}

\begin{document}

\twocolumn[
\begin{@twocolumnfalse}

\maketitle

\begin{center}
\small
*Corresponding authors: 
\href{mailto:mattern@mbi-berlin.de}{mattern@mbi-berlin.de}; 
\href{mailto:bargheer@uni-potsdam.de}{bargheer@uni-potsdam.de}
\end{center}

\vspace{0.5em}

\begin{abstract}
Controlling the spatial distribution of absorbed optical energy is central to nanoscale photothermal chemistry, plasmonics, and ultrafast materials control.
Here, we show that a metallic Au–Pt superlattice concentrates electronic energy in Pt within a few hundred femtoseconds, regardless of the initial energy distribution between the two constituents.
Ultrafast X-ray diffraction follows this energy redistribution through the amplitude of a coherent \textcolor{black}{570}\,GHz superlattice phonon driven by the stress imbalance at the Au-Pt interfaces.
Despite the nearly homogeneous absorption at 400\,nm, the observed  lattice motion is identical to that produced by 800\,nm excitation, which is absorbed predominantly in Pt.
This dissipation-driven localization of energy arises from the large electronic heat capacity of Pt and rapid electronic transport through the superlattice, providing a route to femtosecond control of nanoscale energy distributions.
\end{abstract}

\vspace{1.5em}

\end{@twocolumnfalse}
]

\section{Introduction}

Laser-excited electrons are central to plasmonic nanoscience\cite{herr2023}, photothermal catalysis~\cite{baum2019,verm2024} and ultrafast processes such as  phase transitions \cite{shi2019,matt2025} and energy conversion processes.~\cite{bloc2019}
The term “hot electrons” is often used to conflate two distinct cases: individual non-thermal carriers far above the Fermi level and thermalized electronic distributions with an elevated temperature.~\cite{dubi2019,baum2019,bloc2019,herr2023,verm2024}
However, the distinction is crucial since the thermalization of the photo-excited electrons and electron-phonon coupling ~\cite{wald2016,carp2006,hohl2000,vasi2018} can result in inhomogeneous temperature profiles within metallic heterostructures as a result of energy redistribution via electrons.\cite{pude2020,matt2022,jare2024,stet2025}
Two metals at the same electron temperature store very different electronic energy densities if their electronic heat capacities differ, often dictated by the proximity of d-band states to the Fermi level~\cite{lin2008}.
Combining coinage metals with metal layers that exhibit large electronic heat capacity and strong electron-phonon coupling tailors this electronic energy transport within metallic heterostructures by selectively localizing the energy within these layers~\cite{choi2014,pude2020,matt2022,shin2020,igar2023,kova2015}, even in bilayers which are only a few nanometers thick~\cite{pude2018}.
The concept works irrespective of the 3D arrangement of the constituent materials into layers, laterally structured samples or bimetallic colloids.\cite{stet2025,pude2020,swea2016,asla2017}
It is crucial to understand whether the energy localization arises from strong local electron-phonon interaction or if it occurs already within the electron system, providing e.g. abundant electronic energy to drive chemical reactions.
This was recently clarified using \textcolor{black}{material-selective} ultrafast hard X-ray diffraction (UXRD), which is sensitive to energy rather than temperature.\cite{pude2026}
Bragg peak shifts quantify lattice expansion driven by electronic and phononic stresses via material- and subsystem-specific Grüneisen parameters\cite{matt2023a}.
Since these parameters are often comparable, UXRD largely probes the total local energy density rather than the electron or phonon temperatures.
This distinguishes UXRD from electron-sensitive probes such as transient reflectivity and photoemission, whose signals depend strongly on nonequilibrium electron relaxation and transport\cite{scho1987,reth2002,beya2020}.
\textcolor{black}{The layer-specifc strain response has been used to study the flow of energy within a layer\cite{cava2000,reis2001,sond2008,nico2011,isse2013,matt2021,nguy2026} and across interfaces in a heterostructure\cite{pude2018,matt2022,herz2022,sri2022}.
Importantly, the depth sensitivity of UXRD is not limited to the surface region, but also allows direct probing of energy transport into buried detection layers via hot electrons~\cite{pude2018,pude2020,jare2026}}.
Thinner layers benefit the time-resolution\cite{shay2022}, but suffer from weak UXRD signals. 
Superlattices (SLs) overcome this limitation by combining ultrathin layers with a large scattering volume, yielding intense SL reflections with sub-picosecond temporal sensitivity.
In these structures stress gradients drive backfolded coherent SL phonons whose period is set by the individual bilayer thickness rather than the total stack thickness~\cite{barg2004}.
In extremely well-studied metal-insulator\cite{roth2019} and semiconductor SLs~\cite{yama1994,barg2004,ezza2007,herz2010,boja2012a,ye2025}, such stress gradients are typically imposed by layer-dependent optical absorption.
In metal-metal SLs, electronic pressure can likewise generate large-amplitude coherent THz phonons~\cite{pude2026}, while preserving efficient electronic transport through the entire structure.

\begin{figure}[!tbh]
    \centering
    \includegraphics[width=1\linewidth]{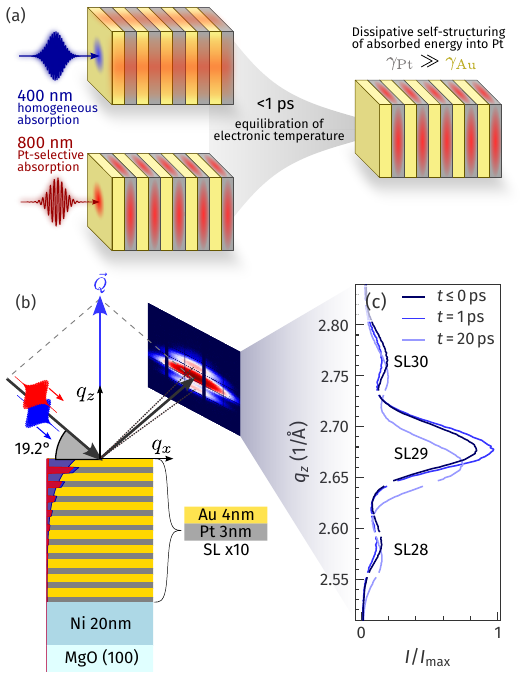}
    \caption{UXRD experiment probes SL phonon.
    (a) Visualization of the energy flow within the SL structure following excitation with either $400\,\text{nm}$ (homogeneous) or $800\,\text{nm}$ (Pt-selective). Red shaded areas indicate regions with a larger energy density.
    (b) Schematic of the SL sample and the geometry of the UXRD experiment.
    The red and blue shaded areas denote the considerably different absorption profiles for $800\,\text{nm}$ and $400\,\text{nm}$, respectively.
    (c) The diffracted intensity along the out-of-plane reciprocal coordinate $q_z$ at three representative delays recorded by an area detector at a fixed diffraction angle in case of $400\,\text{nm}$ excitation.
    Directly after photo-excitation, we observe an intensity modulation of all Bragg peaks originating from the driven SL phonon mode.
    Later, all peaks shift towards smaller $q_z$ as a result of the thermal expansion of the entire SL.
    }
    \label{fig:Figure1}
\end{figure}

Here, we exploit the material-specific sensitivity of UXRD to probe ultrafast electronic energy localization in a metal-metal SL consisting of a few-nanometer-thick Au and Pt layers.
We compare experiments with pump wavelengths at $800\,\text{nm}$ and $400\,\text{nm}$ (see schematic Fig. 1a for illustration):
At $800\,\text{nm}$ the optical energy is absorbed predominantly in Pt.
In contrast, at $400\,\text{nm}$ the absorbed energy is split almost equally between Au and Pt, resulting in a nearly homogeneous initial energy distribution across each bilayer.
UXRD measurements show the same amplitude of the SL phonon even when the energy was evenly distributed by the optical excitation.
This reveals a sub-picosecond energy localization into Pt that establishes a stress imbalance between Au and Pt much faster than the phonon period ensuring its coherent excitation.
Thus, the final nanoscale energy landscape is not determined solely by the optical absorption profile, but can emerge dynamically through ultrafast dissipative equilibration.

\begin{figure}[tbh]
    \centering
    \includegraphics[width=1\linewidth]{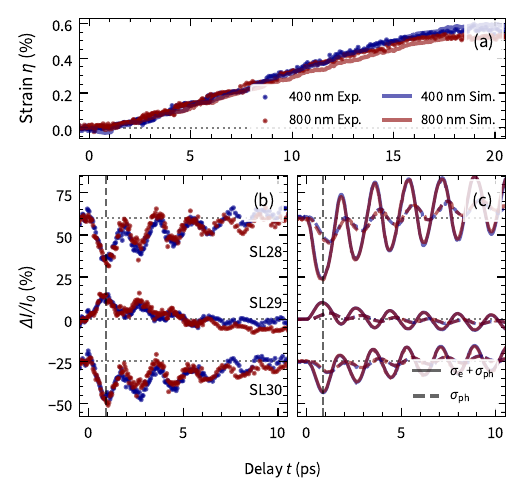}
    \caption{
    Laser-induced dynamics of the Au-Pt SL.
    (a) Average transient strain extracted from the SL Bragg peak shifts, showing the slow expansion of the entire SL structure. The nearly identical maximum strain at $t\approx20\,\text{ps}$ demonstrates similar absorbed fluence for $800$ and $400\,\text{nm}$ excitation.
    (b) Relative intensity changes of the SL Bragg peaks exhibit fast oscillations associated with the \textcolor{black}{$0.57\,\text{THz}$} coherent SL phonon mode. The oscillation amplitudes are nearly identical for both excitation wavelengths.
    (c) Simulations based on a diffusive two-temperature model (d2TM) and the elastic wave equation reproduce the observed dynamics. Including the electronic stress contributions is essential to reproduce the measured oscillation amplitude and phase, whereas phonon stress alone yields a smaller and phase-shifted response. \textcolor{black}{The supplementary information provides additional comparisons, e.g. the oscillations induced by electron stress only.
    }}
    \label{fig:Figure2}
\end{figure}

\section{Experiment and Methods}

We performed UXRD experiments at the FemtoMAX beamline at the MAX IV Laboratory\cite{enqu2018} on an Au-Pt SL consisting of 10 repetitions of $4\,\text{nm}$-thick Au and $3\,\text{nm}$-thick Pt $(111)$-oriented layers grown via molecular beam epitaxy on an MgO substrate with a $20\,\text{nm}$ Ni buffer layer (Figure \ref{fig:Figure1}b).
We studied the laser-induced structural response by tracking the position and the intensity of the SL Bragg peaks on a pixelated position-sensitive area detector in reciprocal space slicing (RSS) geometry\cite{zeus2021}, employing femtosecond hard-X-ray pulses with a photon energy of $8\,\text{keV}$ incident under a fixed angle of $\theta=19.2^{\circ}$ with respect to the sample surface, which probe the entire layer stack.
The mosaicity of the sample reflected by the large width of the Bragg peaks along the in-plane reciprocal coordinate (see Fig.~\ref{fig:Figure1}b) prevents any cross-talk of peak shift and intensity change.

The out-of-plane diffraction profile (Fig. 1c) displays several equidistant SL Bragg peaks - the 28th, 29th and 30th diffraction orders -, whose positions $q_{z}=\frac{2\pi \cdot n}{d_\text{Au}+d_\text{Pt}}$ are set by the Au–Pt bilayer thickness and whose intensities are governed by the thin-layer scattering envelopes\cite{herz2012,barg2004}.

The sample is excited by either $400\,\text{nm}$ or $800\,\text{nm}$ laser pulses with a duration of $50\,\text{fs}$ at a repetition rate of $10\,\text{Hz}$, incident at $26^{\circ}$ with respect to the sample surface.
For $800\,\text{nm}$ we set the incident fluence to $15\,\text{mJ/cm}^{2}$ and adjust the fluence for $400\,\text{nm}$ to the same absorbed fluence, by calibrating to the average thermal expansion long after the pump pulse.
\textcolor{black}{This ensures equal amounts of energy are deposited in both cases, though with significantly different spatial distributions given by the absorption profiles.}
Temporal jitter between the laser and x-ray pulses is minimized using a cross-correlator-based timing tool, providing a timing stability down to $30\,\text{fs}$ within the first picoseconds~\cite{kroo2025}.
Upon laser-excitation, we observe a rapid modulation of the Bragg peaks' intensities and a slowly rising shift of all three Bragg peaks to smaller $q_z$ values (see Fig.~\ref{fig:Figure1}c).
The latter corresponds to an expansion of the entire SL structure that is quantified by the transient strain $\eta(t)~\approx~\left (q_z(t<0\,\text{ps})-q_z(t)\right )/q_{z}(t)$, i.e. the relative change of the SL Bragg peak positions $q_z(t)$ with respect to the position before excitation $q_z(t<0)$ determined via a center-of-mass analysis.
We extract the transient peak intensities by integrating the diffracted intensity over a region of interest that dynamically follows the peak positions to reduce artifacts from peak shifts in the transient intensity.

\begin{figure}[tbh]
    \centering
    \includegraphics[width=1\linewidth]{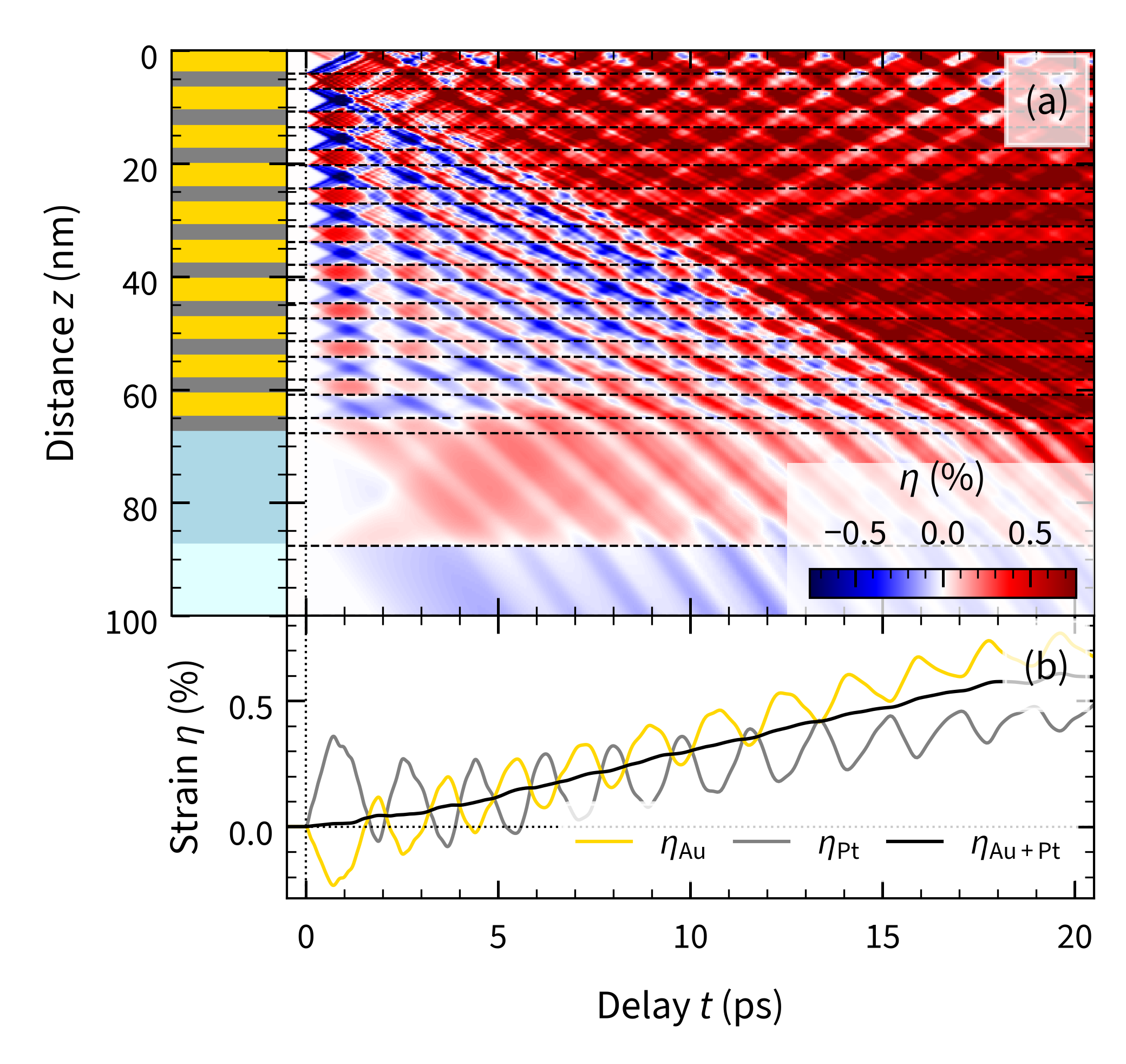}
    \caption{
    Modeled strain dynamics within the Au-Pt SL sample.
    (a) The spatio-temporal strain simulation $\eta(t,z)$ obtained from solving the elastic wave equation shows the opposing strains in Au and Pt associated with the coherent SL phonon mode directly after excitation.
    On longer timescales, the entire SL structure undergoes a thermal expansion that reaches its maximum at $\approx 20\,\text{ps}$, when the acoustic phonon wavepacket has propagated through the SL stack at the speed of sound.
    (b) Averaged strain $\eta(t)$ within the Au and Pt layers as well as the average strain in the entire SL structure.
    The oscillating strain contributions from Au and Pt cancel each other so that the overall strain, measured experimentally as the SL Bragg peak shift, only exhibits a slow rise.
    In contrast, the SL Bragg peak amplitude change provides access to the oscillating strains in the Au and Pt layers and thus shows significantly faster dynamics.
    }
    \label{fig:Figure3}
\end{figure}

\section{Results and Discussion}

The late-time SL Bragg-peak shift calibrates the total absorbed energy, whereas the early-time Bragg-peak intensity oscillation measures the stress imbalance within the Au-Pt bilayers. Thus, comparing these two observables separates total absorbed fluence from nanoscale energy localization.
\begin{figure}[!htbp]
    \centering
    \includegraphics[width=0.9\linewidth]{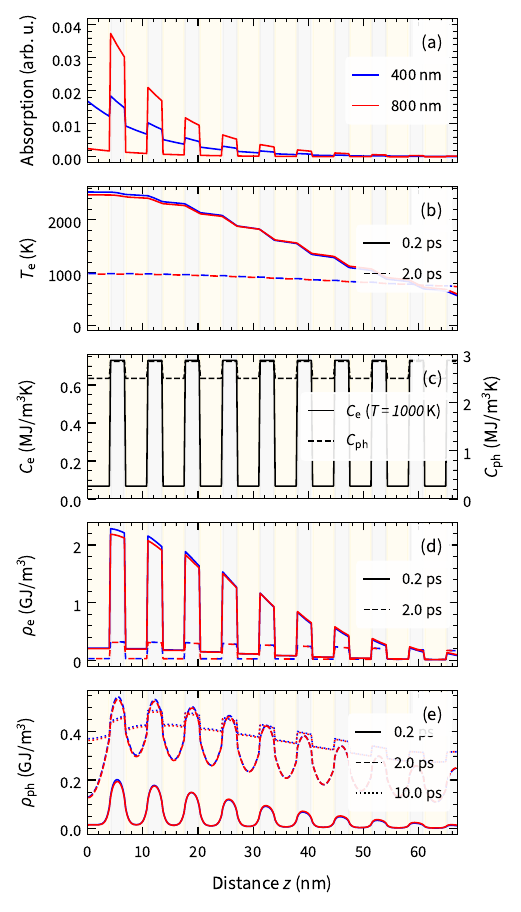}
    \caption{
    Mechanism of dissipative self-structuring of laser-deposited energy.
    (a) Calculated absorption profiles in the Au-Pt SL for $400$ and $800\,\text{nm}$ excitation based on the wavelength-dependent complex refractive indices of Au and Pt.
    While 400\,nm excitation deposits energy nearly homogeneously across Au and Pt, 800\,nm excitation predominantly heats Pt.
    (b) The electron temperature rapidly equilibrates across the bilayers within the first few hundred femtoseconds and over the entire SL stack within the first picoseconds. Note that curves for 400 and 800 nm excitation are identical at 2 ps.
    (c) Spatial modulation of the electronic and phononic heat capacities within the SL structure for the equilibrated electron temperature of approx. $T_e=1000$\,K (cf. panel (b)).
    (d) Due to the large electronic heat capacity of Pt, most of the absorbed electronic energy localizes in the Pt layers after equilibration of the electron temperature. This localization is dictated only by the spatial structuring of the electronic heat capacities (c) and is independent of the initial absorption profile.
    (e) The phononic energy density develops a similar spatial modulation on slower, picosecond timescales due to stronger electron-phonon coupling in Pt. At approximately equal phonon temperatures, the modulation of the phononic energy density remaining at $10\,\text{ps}$ reflects the different volumetric heat capacities of Au and Pt.
    }
    \label{fig:Figure4}
\end{figure}
In detail, the almost identical strain response at large delays ($>10\,\text{ps}$), upon $400$ and $800\,\text{nm}$ excitation (see Fig.~\ref{fig:Figure2}(a)) demonstrates the same absorbed fluences in the SL for both excitation conditions.
The slowly rising strain response reaches a maximum at $20\,\text{ps}$ when the expansion launched at the surface has propagated through the entire SL stack at the speed of sound and has reached the SL-Ni interface (compare strain simulation in Figure~\ref{fig:Figure3}).
In contrast, the peak amplitude change $\Delta I/I_0$ exhibits a significantly faster oscillatory dynamics (see Fig.~\ref{fig:Figure2}b) dictated by the sound velocity $v_\text{s}$ and thickness $d$ of only a single Au-Pt bilayer with a frequency:
 $   f_\text{SL} = \left( \frac{d_\text{Au}}{v_\text{s,Au}} + \frac{d_\text{Pt}}{v_\text{s,Pt}} \right)^{-1} \approx \textcolor{black}{570\,\text{GHz}}$ \textcolor{black}{(see Fig.~S3 for Fourier transformation)}.
This intensity oscillation is a signature of the backfolded coherent SL phonon mode characterized by an expansion of the Pt layers and a concomitant compression of the Au layers.
These opposing strains modulate the SL peak intensities because they change the structure factor of the artificial superlattice unit cell~\cite{herz2012}.
The bilayer period stays nearly constant according to the slow shift of the Bragg peak in Fig. 2a, i.e. the layer strains compensate each other. Hence the peaks only shift on the timescale of the expansion of the entire SL stack~\cite{herz2012} (comp. Figure~\ref{fig:Figure3}).
The opposing strains in the bilayers associated with the coherent SL mode are efficiently driven by the stress imbalance between Au and Pt, provided that this imbalance is established on a timescale shorter than the phonon period.  
Most importantly, we observe identical SL phonon mode responses after excitation with $400\,\text{nm}$ and $800\,\text{nm}$ pump pulses, even though the energy is initially distributed nearly homogeneously across the SL stack at $400\,\text{nm}$ (see red and blue shaded areas in Fig.~\ref{fig:Figure1}b), while 800\,\text{nm} excitation strongly favors absorption in Pt.
These observations already demonstrate that regardless of where the energy is deposited initially, it restructures itself on an ultrafast timescale into the same periodic pattern, which is dictated by the SL structure.
\textcolor{black}{The identical phase and amplitude therefore indicate that a similar ultrafast modulation of the electronic energy density is established for both excitation conditions.}

In order to provide further insights into the mechanisms governing this self-structuring of the energy, we model the laser-induced dynamics using the modular Python library \textsc{udkm1Dsim}~\cite{schi2021}.
In the first step, we reproduce the diffracted intensity before excitation via dynamical X-ray scattering calculations by slightly adjusting the lattice constants of Pt and Au in order to account for growth-induced strains, which influence the static positions and amplitudes of the Bragg peaks.
In the second step, we calculate the optical absorption profile in the framework of an optical transfer matrix algorithm and the subsequent spatial energy redistribution in the framework of a diffusive two-temperature model (d2TM)~\cite{matt2023a,anis1974} using literature values for the thermophysical parameters \textcolor{black}{collected in table~S1 of the supplementary information}.
\textcolor{black}{We also present the simulated lattice dynamics for the limiting cases of a quasi-instantaneous equilibration of electrons and vanishing electronic energy transport in Fig.~S2.
This highlights the crucial role of rapid electronic energy transport for the energy localization in Pt, which is robust against details of the transport.
We therefore neglect additional interfacial thermal resistances in order to avoid introducing poorly constrained fitting parameters.
Reducing the electronic energy transport would furthermore decrease the agreement between the simulated and measured strain in Fig.~\ref{fig:Figure2}a.
}
The resulting spatio-temporal strain response (Figure~\ref{fig:Figure3}) within the heterostructure is then calculated by solving the one-dimensional linear elastic wave equation in the framework of a chain of masses and springs.
In the last step, we calculate the corresponding time-dependent Bragg peaks via dynamical X-ray scattering and extract the transient intensity and position of the SL Bragg peaks as in the experiment. 
\textcolor{black}{ For more details of the modeling and a collection of all relevant parameters we refer to the supplementary information.}
The simulation reproduces both the strain response and the phase and frequency of the intensity oscillations of the Bragg peaks (solid lines in Figure \ref{fig:Figure2}a and c). 
Importantly, we reproduce the identical amplitude response for $400\,\text{nm}$ and $800\,\text{nm}$ excitation.
The dashed lines in Figure \ref{fig:Figure2}c show the simulated coherent phonon response assuming incoherent phonon stress alone.
The phonon stress is established too slowly compared to the period of the SL phonon and thus provides only a small amplitude response with a phase shift compared to the experiment.
Therefore, the dominant driver behind the coherent SL phonon is the electronic stress as only the combined electronic and phononic stress contributions reproduce the correct amplitude and phase of the SL phonon (indicated by the dashed vertical line at the first extrema in Fig. 2b and c), very similar to the observations in a Cu-Pt SL\cite{pude2026}.
The stronger damping of the intensity oscillation in the experiment might originate from a dephasing caused by layer thickness variations across the probed volume, which leads to slightly different oscillation frequencies, or phonon-phonon scattering which is not included in the simulation.

The initial distribution of the absorbed energy follows the absorption profiles shown in Fig.~\ref{fig:Figure4}a, which depend on the complex-valued wavelength-dependent refractive indices of Au and Pt.
At $800\,\text{nm}$, absorption in the Pt layers dominates, while at $400\,\text{nm}$ the excitation is much more uniform and thus the total absorbed energy is split almost evenly between 4 nm Au and 3 nm Pt.
Within the first few hundred femtoseconds, the temperature of the excited electronic system then equilibrates across the individual bilayers, independent of the chosen excitation wavelength (Fig.~\ref{fig:Figure4}b).
As the electron temperature $T_\text{e}$ equilibrates, the energy is redistributed according to the nanostructured variation of the electronic heat capacity in each layer, determined by the density of states at the Fermi level in Au and Pt (Fig.~\ref{fig:Figure4}c).
An equal $T_\text{e}$ across materials with different heat capacities results in different electronic energy densities:
\begin{align}
\rho_e(T_e)=\int_0^{T_e} C_e(T)\,\mathrm{d}T \approx \frac{1}{2}\gamma T_e^2.
\end{align}
Thus, equilibration of $T_\text{e}$ effectively leads to a localization of the energy density within the layers of larger heat capacity given by the product of the Sommerfeld constant $\gamma$ and the electronic temperature: $C_\text{e} = \gamma \cdot T_\text{e}$.
In our case, the $10$ times higher electronic heat capacity in Pt compared to Au thus results in the rapid redistribution of most of the absorbed energy into Pt (Fig.~\ref{fig:Figure4}d).

The spatio-temporal energy density of the electrons $\rho_\text{e}$ and phonons $\rho_\text{ph}$ translate linearly into lattice stress: 
\begin{align}
\sigma = \Gamma_\text{e}\cdot\rho_\text{e} + \Gamma_\text{ph}\cdot\rho_\text{ph},
\end{align}
via the subsystem- and material-specific Gr\"uneisen parameters $\Gamma_\text{e}$ and $\Gamma_\text{ph}$.
Since the electronic Gr\"uneisen parameter in Au\cite{nico2011} is only $1.5$ times larger than in Pt\cite{jare2024}, the $10$ times larger energy deposited in Pt results in an unbalanced stress at the interface driving the SL phonon mode.
On a picosecond timescale, the energy is predominantly transferred to the phonons in Pt via the strong electron-phonon coupling (Fig.~\ref{fig:Figure4}e), which results in an additional, small contribution to the stress gradient at the interfaces, which alone would drive the coherent SL phonon only with small amplitude (dashed lines in Fig.~\ref{fig:Figure2}c).
Only after approximately $10\,\text{ps}$ (Fig. 4 e), do the phonon temperatures of the Au and Pt layers equilibrate, as previously observed for a Au-Ni bilayer\cite{pude2018}.

Due to the spontaneous localization of the absorbed energy in the Pt layers, the SL mode is always driven by the ultrafast expansion of Pt, which results in identical oscillation amplitudes and phases, independent of the spatial structuring of the energy directly after excitation.
\textcolor{black}{We emphasize that these observations are characteristic for heterostructures composed of two metals with considerably different density of states at the Fermi level. In semiconductor heterostructures band structure engineering of two materials with overlapping conduction band but very different effective masses may yield similar effects.}

\section{Conclusion}

In conclusion, we have shown that the coherent \textcolor{black}{$570\,\text{GHz}$} SL phonon mode in a Au-Pt SL has the same amplitude for both homogeneous ($400\,\text{nm}$) and Pt-selective ($800\,\text{nm}$) excitation of the bilayer for the same  \textcolor{black}{total absorbed energy, which is uniquely measured by the Bragg peak shift. }
The large electronic heat capacity of Pt localizes most of the optically deposited energy, resulting in an unbalanced stress at the interface of the Au-Pt bilayers in our experiment, driving the coherent SL mode even when Au and Pt absorb nearly equal energy.
This observation shows that the absorbed energy can rapidly localize on nanometer length- and femtosecond timescales by dissipative equilibration within the electron gas, regardless of the initial energy distribution. Such ultrafast dissipative energy localization may provide a general pathway for engineering transient nanoscale energy hotspots\cite{stet2025,swea2016,asla2017} in photonic, catalytic, and optically driven quantum materials.
Combined with their established use as transducers for THz phonons, SLs therefore form an interesting and versatile class of metamaterials for ferrimagnetic and antiferromagnetic spintronics\cite{kim2022,dal2024,jung2016} and for future experiments providing unique insights into the electronic energy transport within metallic heterostructures.

\section*{Data Availability}
The data and simulation scripts of this study will be openly available in Zenodo. 

\section*{Notes} The authors declare no competing financial interest.

\section*{Acknowledgments}
J.J., M.M., and D.S. would like to thank the Leibniz Association for funding through the Leibniz Junior Research Group J134/2022.
L.M., M.H. and M.B. acknowledge funding from the Deutsche Forschungsgemeinschaft (DFG, German Research Foundation) – CRC/SFB 1636 – Project ID 510943930 – Project No. A01 and Z02. S.Z. and M.B. acknowledge support by the BMBF via project No. 05K22IP1.
C.W. and M.B. acknowledge the DFG for financial support for Project No. 328545488 — TRR 227, project A10. M.K. was funded by the DFG Project 422314695032 - CRC/SFB 1277 Project B12.
J.-E.P. would like to thank the European X-ray Free-Electron Laser Facility for funding. J.L. acknowledges support from the Swedish Research Council (VR, Grant No. 2023-05136) and the Olle Engkvists Stiftelse (Grant No. 238-0012).
We acknowledge the MAX IV Laboratory for beamtime on the FemtoMAX beamline under Proposal No. 20240531.
Research conducted at MAX IV, a Swedish national user facility, is supported by Vetenskapsrådet (Swedish Research Council, VR) under Contract No. 2018-07152, Vinnova (Swedish Governmental Agency for Innovation Systems) under Contract No. 2018-04969, and Formas under Contract No. 2019-02496. We acknowledge
Dr. Matthias Rössle for ellipsometric characterization
of the SL as well as Pt and Au layers.
\bibliographystyle{achemso}
\bibliography{references.bib}

\end{document}